\documentclass[10pt,twocolumn]{article}

\usepackage[
    letterpaper,
    top=0.65in,
    bottom=0.68in,
    left=0.67in,
    right=0.67in
]{geometry}

\usepackage{amsmath}
\usepackage{booktabs}
\usepackage{graphicx}
\usepackage{microtype}
\usepackage{enumitem}
\usepackage[numbers,sort&compress]{natbib}
\usepackage{xurl}
\usepackage[hidelinks]{hyperref}
\usepackage{tikz}
\usepackage{titlesec}

\usetikzlibrary{arrows.meta,positioning}

\setlist{
    nosep,
    leftmargin=*
}

\titlespacing*{\section}
{0pt}
{1.05ex plus 0.25ex minus 0.15ex}
{0.45ex}

\titlespacing*{\subsection}
{0pt}
{0.80ex plus 0.20ex minus 0.10ex}
{0.30ex}

\begin{document}


\twocolumn[
\begin{center}

{\fontsize{16.5}{18.5}\selectfont\bfseries
FoldPipe: Bounded Remote Streaming of Native Molecular Shards\\[-0.05em]
with Asynchronous Prefetch
\par}

\vspace{0.38em}

{\large Dhiren Mukesh Khatri\par}
{\normalsize Independent Researcher\par}

\vspace{0.62em}

\begin{minipage}{0.94\textwidth}
\small

\textbf{Abstract---}
Training molecular machine-learning models on ephemeral or
memory-constrained accelerator instances can require repeatedly retrieving
preprocessed molecular graphs from remote storage.
FoldPipe is a lightweight Python orchestration layer for already-sharded
PyTorch and PyTorch Geometric data.
It retrieves one shard ahead in a background thread while the consumer
trains on the current shard, keeping the number of live shard payloads
bounded with respect to total dataset size.

Asynchronous prefetch and bounded buffering are established systems
techniques rather than novel scheduling algorithms.
FoldPipe's contribution is a small integration targeted at native
\texttt{.pt} molecular shards together with a source-pinned empirical
characterization of its operating regime.

We evaluate a SchNet energy-and-force workload on MD17 aspirin using
20 paired, order-alternating benchmark passes on a Tesla T4.
Each pass processes five pinned shards containing 25,000 structures.
FoldPipe records 16.33~s mean I/O--compute overlap, compared with zero by
construction for the sequential bounded baseline.
Mean pass time is 76.78~s for FoldPipe and 83.37~s for the baseline.
However, the geometric mean paired speedup is
1.059$\times$ with a 95\% bootstrap interval of
0.878$\times$--1.288$\times$.
The experiment therefore verifies the overlap mechanism but is
inconclusive about a reliable wall-clock speed advantage under the
observed public-network variability.

\end{minipage}

\vspace{0.65em}

\end{center}
]

\section{Introduction}

Molecular machine-learning workloads commonly represent molecular
conformations as graphs containing atomic identities, coordinates,
energies, and forces.
Architectures such as SchNet~\cite{schutt2018schnet} can operate on these
representations through PyTorch~\cite{paszke2019pytorch} and
PyTorch Geometric (PyG)~\cite{fey2019pyg}.

In temporary notebook or cloud environments, however, processed training
data may reside in remote storage while host memory and persistent disk are
limited.
A researcher may already possess hundreds or thousands of serialized
PyTorch or PyG shard files and may not want to migrate them into a new
database or archive format merely to stream them to a training process.

FoldPipe addresses this narrow integration problem.
It consumes existing bounded-size \texttt{.pt} shards through a small
source interface and overlaps retrieval of shard $i+1$ with computation on
shard $i$.

The system deliberately does not claim that producer--consumer queues,
bounded buffering, asynchronous prefetch, or the resulting pipeline
equations are new.
Prefetch and computation--I/O overlap are long-established systems
techniques~\cite{shriver1999prefetching}.
The contribution is instead the application boundary: direct remote
consumption of native PyTorch/PyG molecular shards together with an
instrumented evaluation that distinguishes mechanism from observed
end-to-end speed.

This paper makes three contributions:

\begin{enumerate}
    \item a small backend-independent interface for remotely stored native
    PyTorch/PyG shards;

    \item a one-stage asynchronous pipeline whose number of live shard
    payloads remains constant with respect to total shard count;

    \item a paired MD17/SchNet experiment that directly measures retrieval,
    deserialization, computation, overlap, memory, and runtime while
    reporting uncertainty rather than assuming that overlap necessarily
    produces a speedup.
\end{enumerate}

\section{Related Work}

PyTorch provides general-purpose data loading and multiprocessing
facilities~\cite{paszke2019pytorch}.
PyG provides \texttt{OnDiskDataset} for graph objects that do not fit in
CPU memory and \texttt{PrefetchLoader} for asynchronously moving prepared
batches toward the accelerator~\cite{pygondiskdocs,pygloaderdocs}.
FoldPipe operates at an earlier boundary: retrieval and deserialization of
remote shard files before ordinary PyG batching.

WebDataset packages samples into tar shards designed for sequential local
or remote streaming~\cite{aizman2020webdataset}.
The Hugging Face Hub provides revision-addressable repository storage and
remote file retrieval~\cite{huggingfacehub}.
FoldPipe does not attempt to replace either system.
Its narrower compatibility advantage is that already-generated
\texttt{.pt} shards can be consumed without first being converted into tar
records or a database.

Broader input-pipeline systems such as \texttt{tf.data} incorporate
parallelism, caching, transformations, and prefetch at considerably larger
scope~\cite{murray2021tfdata}.
FoldPipe intentionally remains a small orchestration layer rather than a
general input-processing framework.

\section{FoldPipe Design}

\subsection{Source abstraction}

A FoldPipe source exposes two conceptual operations:

\begin{enumerate}
    \item lazily enumerate shard identifiers; and
    \item retrieve and deserialize one shard.
\end{enumerate}

Current software supports Hugging Face and Google Drive sources, together
with synthetic and pre-enumerated sources used in testing and controlled
experiments.

The Hugging Face backend can pin retrieval to an exact repository revision.
Remote bytes are streamed into memory and deserialized on the CPU.
Consequently, FoldPipe avoids a required local-disk cache, but it is not
zero-copy: serialized bytes and deserialized objects may coexist briefly.

Because native PyTorch/PyG objects are deserialized using
\texttt{torch.load(weights\_only=False)}, shard files must be trusted.
FoldPipe is not intended to safely execute arbitrary untrusted serialized
objects.

\subsection{One-stage asynchronous prefetch}

The core iterator uses a single background worker.
After shard $i$ becomes available, retrieval of shard $i+1$ is submitted
before batches from shard $i$ are yielded to the consumer.

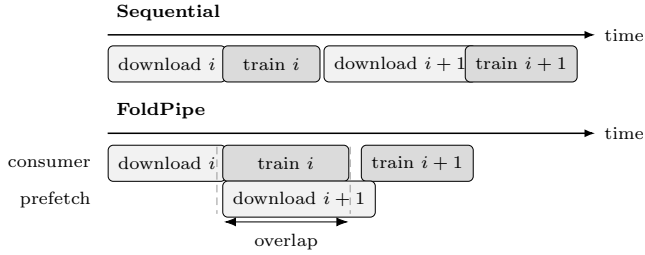
\begin{figure}[t]
\centering

\resizebox{0.98\columnwidth}{!}{%
\begin{tikzpicture}[
    x=0.74cm,
    y=0.62cm,
    font=\scriptsize,
    io/.style={
        draw,
        rounded corners=2pt,
        fill=black!5,
        minimum height=0.50cm,
        align=center,
        inner xsep=4pt
    },
    compute/.style={
        draw,
        rounded corners=2pt,
        fill=black!14,
        minimum height=0.50cm,
        align=center,
        inner xsep=4pt
    },
    sectionlabel/.style={
        font=\scriptsize\bfseries,
        anchor=west
    },
    rowlabel/.style={
        font=\scriptsize,
        anchor=east
    },
    timeline/.style={
        -{Latex[length=1.5mm]},
        semithick
    }
]


\node[sectionlabel] at (0,4.15) {Sequential};

\draw[timeline]
    (0,3.65) -- (9.15,3.65)
    node[right] {\scriptsize time};

\node[
    io,
    minimum width=1.55cm,
    anchor=west
] (seq-d1) at (0,3.00)
{download $i$};

\node[
    compute,
    minimum width=1.35cm,
    anchor=west
] (seq-c1) at (2.15,3.00)
{train $i$};

\node[
    io,
    minimum width=1.75cm,
    anchor=west
] (seq-d2) at (4.05,3.00)
{download $i+1$};

\node[
    compute,
    minimum width=1.55cm,
    anchor=west
] (seq-c2) at (6.70,3.00)
{train $i+1$};


\node[
    sectionlabel
] at (0,1.95) {FoldPipe};

\draw[timeline]
    (0,1.45) -- (9.15,1.45)
    node[right] {\scriptsize time};

\node[rowlabel] at (-0.15,0.78) {consumer};

\node[
    io,
    minimum width=1.55cm,
    anchor=west
] (fp-d1) at (0,0.78)
{download $i$};

\node[
    compute,
    minimum width=1.75cm,
    anchor=west
] (fp-c1) at (2.15,0.78)
{train $i$};

\node[
    compute,
    minimum width=1.55cm,
    anchor=west
] (fp-c2) at (4.75,0.78)
{train $i+1$};

\node[rowlabel] at (-0.15,-0.02) {prefetch};

\node[
    io,
    minimum width=1.75cm,
    anchor=west
] (fp-d2) at (2.15,-0.02)
{download $i+1$};

\draw[densely dashed,black!40]
    (2.05,1.15) -- (2.05,-0.35);

\draw[densely dashed,black!40]
    (4.55,1.15) -- (4.55,-0.35);

\draw[
    <->,
    >=Latex,
    thin
]
    (2.18,-0.53) -- (4.52,-0.53)
    node[midway,below=1pt] {\scriptsize overlap};

\end{tikzpicture}%
}

\caption{
Sequential streaming performs retrieval and computation one after another.
FoldPipe retrieves shard $i+1$ in the background while shard $i$ is being
consumed, so part of the retrieval latency can be hidden.
}

\label{fig:architecture}
\end{figure}

Figure~\ref{fig:architecture} illustrates the distinction.


Let $D_i$ denote retrieval plus deserialization time for shard $i$, and let
$C_i$ denote time spent consuming it.
Ignoring small orchestration overheads, bounded sequential streaming takes

\begin{equation}
T_{\mathrm{seq}}
=
\sum_{i=1}^{n}
\left(D_i+C_i\right).
\end{equation}

An ideal one-stage pipeline takes approximately

\begin{equation}
T_{\mathrm{pipe}}
=
D_1
+
\sum_{i=1}^{n-1}
\max\left(C_i,D_{i+1}\right)
+
C_n.
\end{equation}

For a homogeneous long stream this approaches

\begin{equation}
\frac{T_{\mathrm{seq}}}{n}
\approx D+C,
\end{equation}

whereas

\begin{equation}
\frac{T_{\mathrm{pipe}}}{n}
\approx \max(D,C).
\end{equation}

Thus, in the ideal matched-latency case $D \approx C$, throughput can
approach a factor-of-two improvement.
If retrieval is much slower than computation, only the computation duration
can be hidden.

These are standard pipeline properties rather than a FoldPipe-specific
algorithmic result~\cite{shriver1999prefetching}.

\subsection{Memory bound}

FoldPipe does not retain every shard in the dataset.
At steady state the consumer may hold the current deserialized shard while
the background future holds the next shard or its in-progress byte buffer.

If individual shards are bounded by $S_{\max}$, the live data-path memory is

\begin{equation}
O(S_{\max})
\end{equation}

with respect to total shard count.
The constant factor can approach two shard payloads, plus mini-batch and
model memory.

This statement does not imply that FoldPipe can never run out of memory.
A single individually oversized shard can still exceed available RAM.

\section{Experimental Method}

\subsection{Workload}

The real-workload experiment uses the aspirin trajectory from
MD17~\cite{chmiela2017md17}, transformed into PyG graph objects.

Five benchmark shards are used.
Each contains 5,000 structures, for 25,000 structures per timed pass.

The dataset repository is pinned to revision
\texttt{f779686deb9217877dd7ddde99b2522bd441492a}.

The workload uses PyG SchNet with 128 hidden channels, 128 filters, six
interaction blocks, 50 Gaussian basis functions, and a 10~\AA{} cutoff.

Batches contain 32 molecular graphs.
Each timed optimization step predicts energy and derives forces from the
negative coordinate gradient.
The loss combines energy mean-squared error with ten times force
mean-squared error.

The experiment ran on a Tesla T4.
One untimed warm-up batch was executed before measurement.

\subsection{Paired protocol}

Twenty pairs were measured.

Within each pair, the sequential and FoldPipe pipelines processed the same
five shards, used the same model initialization and manual seed, and
alternated execution order between pairs.

Therefore the final experiment contains 40 timed passes and 200 timed shard
traces.

The benchmark records monotonic timestamps for download start, download
finish, deserialization finish, training start, and training finish.

Resident memory and GPU utilization are sampled during each pass.
GPU synchronization bounds the per-shard training interval.

The shard list is fixed before timing, so remote dataset discovery is not
included in the timed comparison.

For every pair, runtime speedup is calculated as

\begin{equation}
R_i
=
\frac{T_{\mathrm{seq},i}}
     {T_{\mathrm{FoldPipe},i}}.
\end{equation}

The paired ratios are summarized with their geometric mean.
A deterministic percentile bootstrap of paired log ratios with 20,000
resamples estimates the 95\% interval.

Absolute paired time differences are also reported.

The benchmark was executed from clean source commit
\texttt{16fdbb26b00f9721ce4034335ce0ee12bda77720}.
The corresponding source manifest and source-bundle SHA-256 are preserved
with the public research artifact~\cite{khatri2026benchmark}.

\section{Results}

\begin{table}[t]
\centering

\caption{Twenty-pair MD17/SchNet benchmark summary.}

\label{tab:results}

\small

\begin{tabular}{lrr}
\toprule
Metric & Sequential & FoldPipe \\
\midrule

Mean pass time (s)
& 83.37 & 76.78 \\

Median pass time (s)
& 75.27 & 75.97 \\

Time SD (s)
& 39.86 & 28.46 \\

GPU utilization (\%)
& 36.17 & 37.70 \\

Peak RSS (GiB)
& 1.935 & 2.064 \\

I/O--compute overlap (s)
& 0.00 & 16.33 \\

GPU wait time (s)
& 56.50 & 48.65 \\

\bottomrule
\end{tabular}

\end{table}

FoldPipe was faster in 11 of 20 pairs.

The geometric mean paired runtime ratio was
$1.059\times$, with a 95\% paired-bootstrap interval of
$0.878\times$--$1.288\times$.

Because this interval includes $1.0$, the experiment does not establish a
reliable wall-clock speed advantage.

Mean paired time saved was 6.59~s with a 95\% interval from
$-7.69$ to 21.11~s.
Median paired time saved was 3.45~s with an interval from
$-13.48$ to 22.89~s.
These intervals also contain the no-effect value.

The mechanism measurement answers a different question.
FoldPipe recorded 16.33~s mean retrieval--computation overlap per pass,
whereas the sequential implementation has no concurrent retrieval and
training by construction.

The experiment therefore provides direct evidence that the implementation
performs the intended overlap, even though variable remote-network
conditions prevent the current sample from establishing a reliable overall
speed advantage.

Peak RSS is slightly higher for FoldPipe because a prefetched next payload
may coexist with the shard currently being consumed.
The experiment therefore does not support a claim that asynchronous
streaming requires less memory than bounded sequential streaming.
Instead, both retain a working set that is independent of total shard count.

\section{Discussion}

The results separate \emph{mechanism} from \emph{outcome}.

FoldPipe successfully overlaps remote retrieval with current-shard
computation.
That fact alone does not guarantee lower end-to-end runtime because public
network latency can vary substantially between consecutive passes.

The simple pipeline model predicts the same boundary.
When $C_i$ is comparable to $D_{i+1}$, a large portion of next-shard
retrieval can be hidden.
When $D_{i+1} \gg C_i$, the consumer must still wait for the remaining
download.

The real MD17 experiment therefore should not be interpreted as evidence
for a universal speedup.
Instead, it shows that the mechanism operates as intended and that its
throughput value depends on the relationship between remote I/O latency and
per-shard computation.

This positioning also distinguishes FoldPipe from more comprehensive input
systems.
Researchers willing to migrate their data to WebDataset, a database-backed
PyG dataset, or another specialized representation may obtain functionality
that FoldPipe does not provide.
FoldPipe is most useful when existing native \texttt{.pt} shards and minimal
format migration are primary constraints.

\section{Limitations}

The evaluation has several limitations.

First, the principal real-workload benchmark uses a public remote network.
Alternating pipeline order reduces a simple ordering bias but does not make
the two pipelines experience identical network conditions.

Second, twenty pairs still produce wide uncertainty intervals.
Larger experiments would provide tighter estimates of the throughput
effect.

Third, the real workload uses one molecular system, one model family, and
one accelerator configuration.
The results therefore characterize this workload rather than all molecular
training pipelines.

Fourth, FoldPipe operates at shard granularity.
It does not provide sample-level remote streaming inside a shard, automatic
distributed sharding, generalized transformation graphs, or the storage
features of mature database and archive systems.

Finally, native PyTorch object deserialization is unsafe for untrusted
files.
Only trusted serialized shards should be used.

\section{Software and Data Availability}

The FoldPipe implementation is publicly available as open-source software
under the MIT license.

The current software release is version 0.3.2.
The MD17 measurements reported in this paper are preserved as the frozen
v0.3.0 research benchmark artifact~\cite{khatri2026benchmark}.

The public repository includes the benchmark report, raw statistics,
per-shard traces, source manifest, execution log, benchmark image, and the
scripts used to construct the processed benchmark shards.

The benchmark data are derived from the MD17 aspirin dataset rather than
being introduced as a new molecular dataset in this paper.


\begin{figure*}[!t]
\centering

\includegraphics[
    width=0.70\textwidth
]{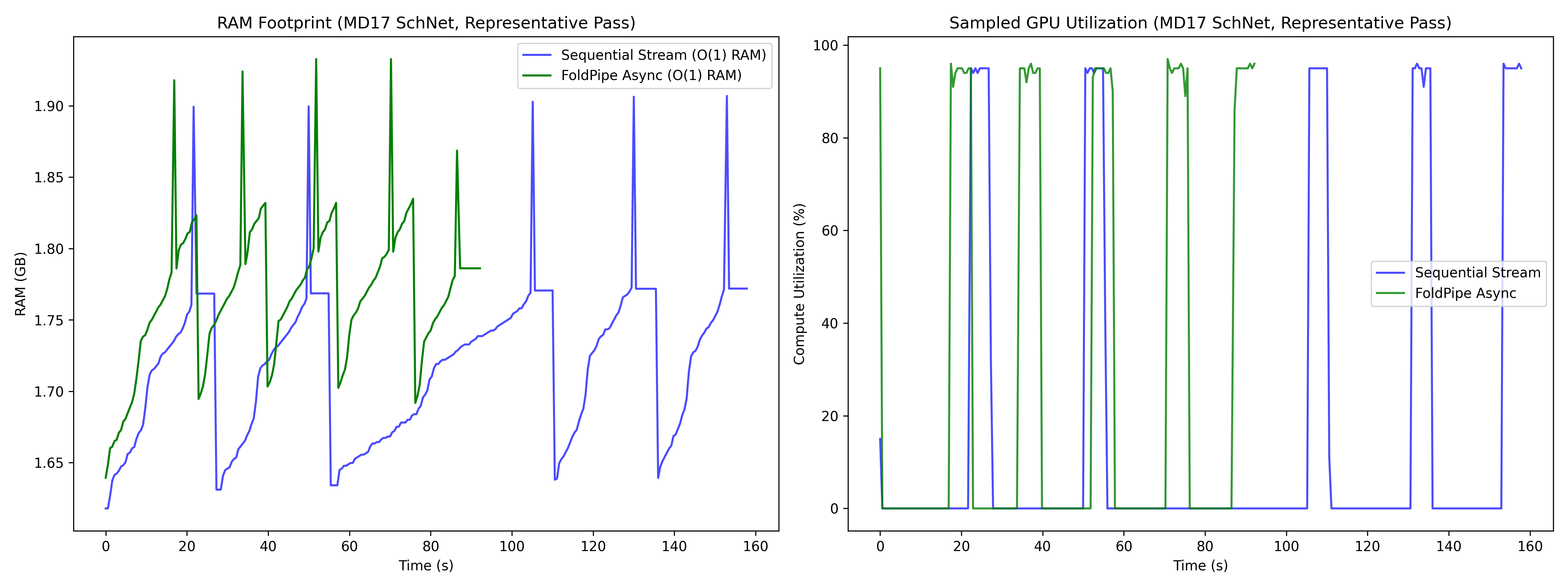}

\vspace{-0.35em}

\caption{
Representative resident-memory and sampled-GPU-utilization traces from the
MD17/SchNet benchmark.
These traces illustrate execution behavior; statistical inference uses all
20 paired runs rather than this individual example.
}

\vspace{-0.45em}

\label{fig:benchmark}
\end{figure*}

\section{Conclusion}

FoldPipe provides a small bounded-working-set streaming layer for existing
native PyTorch and PyG molecular-data shards stored remotely.

Its single-stage prefetcher retrieves the next shard while computation
proceeds on the current shard.
The resulting data-path memory requirement is bounded by shard size and
prefetch depth rather than by total dataset size.

A 20-pair MD17/SchNet experiment directly measured 16.33~s mean
I/O--compute overlap.
Observed mean runtime favored FoldPipe, but paired uncertainty intervals
included the no-effect value.

The present experiment therefore demonstrates the overlap mechanism while
remaining inconclusive about a reliable wall-clock speed advantage under
variable public-network conditions.

FoldPipe should consequently be viewed as a low-conversion integration for
remote native molecular shards whose performance benefit depends on the
I/O--compute regime, rather than as a new prefetch algorithm or a
universally faster replacement for established data systems.


\bibliographystyle{unsrtnat}
\bibliography{references}

\end{document}